\newif\ifColor\Colortrue \Colorfalse
\Colortrue
\PassOptionsToPackage{super,comma,numbers,sort&compress}{natbib}  
\documentclass[preprint, 5p, 8pt, lefttitle]{elsarticle}

\usepackage{xpatch}
\xpatchcmd{\MaketitleBox}{\hrule\vskip12pt}{\vspace{-2\baselineskip}}{}{}% remove first horizontal rule (above abstract)
\xpatchcmd{\MaketitleBox}{\hrule}{}{}{}% remoce second horizonral rule (below keywords)

\usepackage[switch, left]{lineno}
\modulolinenumbers[1]
\usepackage[utf8]{inputenc}
\usepackage{amssymb}
\usepackage{amsmath}
\usepackage{amsfonts}
\usepackage{upgreek}
\usepackage{gensymb}
\usepackage{setspace}
\usepackage{booktabs}
\usepackage[english]{babel}
\usepackage{mathtools}
\usepackage{mhchem}
\usepackage{hyperref}
\usepackage[final]{microtype}
\usepackage{csquotes,lmodern}
\usepackage{subcaption}
\usepackage{etoolbox}
\usepackage{color}
\usepackage{multibib}
\usepackage{xspace}
\newcites{methods}{References}
\usepackage[labelfont=bf]{caption}
\DeclareCaptionLabelSeparator{thinspace}{.\,}
\DeclareCaptionLabelFormat{mylabel}{#1\,#2}

\usepackage{sidecap}
\usepackage{adjustbox}
\newcommand{\PFunit}{mW\,m$^{-1}$\,K$^{-2}$\xspace}
\makeatletter
\renewenvironment{abstract}{\global\setbox\absbox=\vbox\bgroup
  \hsize=\textwidth\def\baselinestretch{0}%
 \par\unskip\noindent\unskip\ignorespaces}
 {\egroup}

\def\keyword{%
  \def\sep{\unskip, }%
 \def\MSC{\@ifnextchar[{\@MSC}{\@MSC[2000]}}
  \def\@MSC[##1]{\par\leavevmode\hbox {\it ##1~MSC:\space}}%
 \def\PACS{\par\leavevmode\hbox {\it PACS:\space}}%
  \def\JEL{\par\leavevmode\hbox {\it JEL:\space}}%
 \global\setbox\keybox=\vbox\bgroup\hsize=\textwidth
  \normalsize\normalfont\def\baselinestretch{0}
 \parskip\z@
  \noindent\textit{Some important words: }   <--- Edit as necessary
  \raggedright                         % Keywords are not justified.
  \ignorespaces}

\def\ps@pprintTitle{%
     \let\@oddhead\@empty
     \let\@evenhead\@empty
     \def\@oddfoot{\footnotesize\itshape
      \ifx\@journal\@empty  % <--- Edit as necessary
       \else\@journal\fi\hfill\today}%
     \let\@evenfoot\@oddfoot}
     
\makeatother

\begin{document}
\begin{frontmatter}

\title{\onehalfspacing\textbf{\fontsize{18}{18}\selectfont{\textsf{%Ultrahigh thermoelectric power factor in \ce{Ni3Ge} by intrinsic energy filtering
Energy filtering-induced ultrahigh thermoelectric power factors in \ce{Ni3Ge}
}}}}

\author[1]{\textsf{\small\textbf{Fabian Garmroudi}}\corref{cor1}}
\author[2]{\textsf{\small\textbf{Simone Di Cataldo}}} 
\author[1]{\textsf{\small\textbf{Michael Parzer}}}
\author[3]{\textsf{\small\textbf{Jennifer Coulter}}}
\author[4]{\textsf{\small\textbf{Yutaka Iwasaki}}}
\author[1]{\textsf{\small\textbf{Matthias Grasser}}}
\author[1]{\textsf{\small\textbf{Simon Stockinger}}}
\author[1]{\textsf{\small\textbf{Stephan Pázmán}}}
\author[1]{\textsf{\small\textbf{Sandra Witzmann}}}
\author[1]{\textsf{\small\textbf{Alexander Riss}}}
\author[1]{\textsf{\small\textbf{Herwig Michor}}}
\author[5]{\textsf{\small\textbf{Raimund Podloucky}}}
\author[6]{\textsf{\small\textbf{Sergii Khmelevskyi}}}
\author[3,7,8,9]{\textsf{\small\textbf{Antoine Georges}}}
\author[1]{\textsf{\small\textbf{Karsten Held}}}
\author[4,10]{\textsf{\small\textbf{Takao Mori}}}
\author[1]{\textsf{\small\textbf{Ernst Bauer}}}
\author[1]{\textsf{\small\textbf{Andrej Pustogow}}\corref{cor1}}
\cortext[cor1]{f.garmroudi@gmx.at, pustogow@ifp.tuwien.ac.at}
\address[1]{\textsf{Institute of Solid State Physics, TU Wien, 1040 Vienna, Austria}}
\address[2]{\textsf{Dipartimento di Fisica, Sapienza University of Rome, Piazzale Aldo Moro 5, 00185 Roma, Italy}}
\address[3]{\textsf{Center for Computational Quantum Physics, Flatiron Institute, New York 10010, USA}}
\address[4]{\textsf{International Center for Materials Nanoarchitectonics (WPI-MANA), National Institute for Materials Science, Tsukuba, Japan}}
\address[5]{\textsf{Institute of Materials Chemistry, Universit\"at Wien, 1090 Vienna, Austria}}
\address[6]{\textsf{Vienna Scientific Cluster Research Center, TU Wien, 1040 Vienna, Austria}}
\address[7]{\textsf{Collège de France, PSL University, 75005 Paris, France}}
\address[8]{\textsf{Department of Quantum Matter Physics, University of Geneva, 1211 Geneva, Switzerland}}
\address[9]{\textsf{Centre de Physique Théorique, Ecole Polytechnique, 91128 Palaiseau, France}}
\address[10]{\textsf{Graduate School of Pure and Applied Sciences, University of Tsukuba, Tsukuba, Japan}}
\selectlanguage{english}
%\journal{submitted to Science Advances}

\begin{abstract}
\noindent \textsf{\fontsize{10}{10}\selectfont{Traditional thermoelectric materials rely on low thermal conductivity to enhance their efficiency but suffer from inherently limited power factors. Novel pathways to optimize electronic transport are thus crucial. Here, we achieve ultrahigh power factors in \ce{Ni3Ge}-based systems through a new materials design principle. When overlapping flat and dispersive bands are engineered to the Fermi level, charge carriers can undergo intense interband scattering, yielding an energy filtering effect similar to what has long been predicted in certain nanostructured materials.
Via a multi-step DFT-based screening method developed herein, we discover a new family of L1$_2$-ordered binary compounds with ultrahigh power factors up to \,11\,\PFunit near room temperature, which are driven by an intrinsic phonon-mediated energy filtering mechanism. Our comprehensive experimental and theoretical study of these new intriguing materials paves the way for understanding and designing high-performance scattering-tuned metallic thermoelectrics.
}}
\end{abstract}
\end{frontmatter}

\section*{INTRODUCTION}
\vspace*{-0.1cm}
%\begin{linenumbers}
\noindent Waste heat is ubiquitous and the majority of it occurs in a decentralized and low-grade form \cite{forman2016estimating}. Thermoelectrics (TEs) present a promising solution for harvesting a small fraction of this heat, for example, to power trillions of autonomous sensor systems and smart devices expected to be installed with the upcoming Internet of Things \cite{pecunia2023roadmap}. Efficient TE materials require a large power factor ($P\!F = S^2\sigma$) and low thermal conductivity ($\kappa$), which together determine the dimensionless figure of merit $zT = S^2\sigma\kappa^{-1}T$; here, $S$, $\sigma$ and $T$ denote the Seebeck coefficient, electrical conductivity and temperature, respectively. Despite significant advancements in the discovery of new TE materials with high $zT$ \cite{zhao2014ultralow,zhao2014high,mao2019high}, integrating these materials into practical applications remains an unresolved key challenge \cite{yan2022high}, and, so far, only \ce{Bi2Te3}-based systems, discovered 70 years ago \cite{goldsmid1954use}, are commercially available. 

A major bottleneck is that many of the current high-performance materials have mechanical and thermal stability issues -- an unfortunate consequence following from the fact that conventional TE materials design focuses on maximizing structural and chemical complexity to restrict lattice-driven heat transport \cite{snyder2008complex,toberer2010zintl}. On the other hand, an optimization approach that focuses on high $zT$ arising from high $P\!F$ rather than from low $\kappa$ will more effectively lead to materials with a real-world impact. Indeed, among all TE materials, robust and stable half-Heusler compounds with high $P\!F$ are currently among prime candidates for making it into practical applications \cite{fey2023isa}, despite their $zT$ being significantly smaller than that of other high-performance TE materials. 

Enhancing $P\!F$, however, is much less straightforward and particularly difficult due to the inherent trade-off between $S$ and $\sigma$ -- arguably the toughest challenge in designing TE materials. Consequently, the power factor of most semiconductors is usually far below 10\,\PFunit, with rare exceptions found in few half- and full-Heusler compounds \cite{he2016achieving,garmroudi2021boosting}. A number of dedicated strategies to overcome this limitation have been developed, e.g., aligning multiple conduction or valence bands (band convergence) \cite{pei2011convergence}, modulation doping \cite{yu2012enhancement}, or even utilizing magnetic interactions \cite{mori2017novel}. Another enhancement concept which has received great interest is energy filtering. It has long been predicted that in certain nanocomposites, electronic scattering could be leveraged to filter out low-energy charge carriers \cite{heremans2005thermopower,faleev2008theory}, but verification of this concept has been an outstanding experimental challenge \cite{gayner2020energy,ghosh2023towards} and exploitation for practical purposes remains severely limited \cite{soni2012interface,masci2024large} up until today. 

\begin{figure}[t!]
\newcommand{\setwidth}{0.45}
			\centering
			\hspace*{0cm}
		\includegraphics[width=0.45\textwidth]{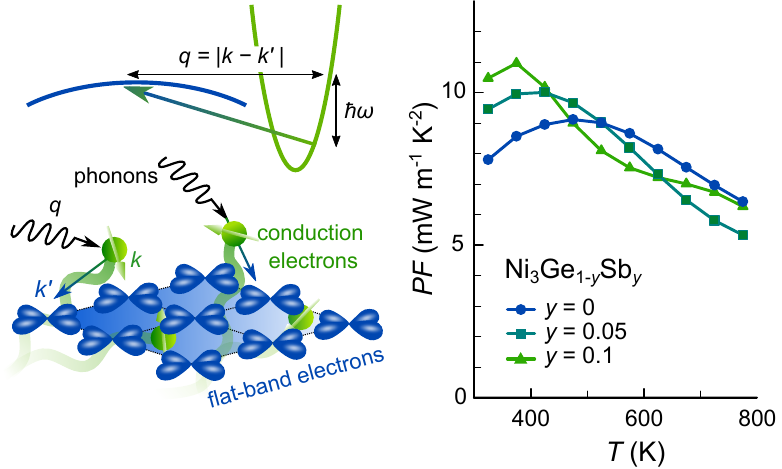}
	\caption{Schematic of phonon-mediated interband scattering and energy filtering of low-energy carriers in metallic \ce{Ni3Ge}, leading to high $P\!F > 10\,$mW\,m$^{-1}$\,K$^{-2}$ ($zT \approx 0.3$) around room temperature.} 
	\label{Fig1}
\end{figure}

\begin{figure*}[tbh]
\newcommand{\setwidth}{0.45}
			\centering
			\hspace*{0cm}
		\includegraphics[width=0.9\textwidth]{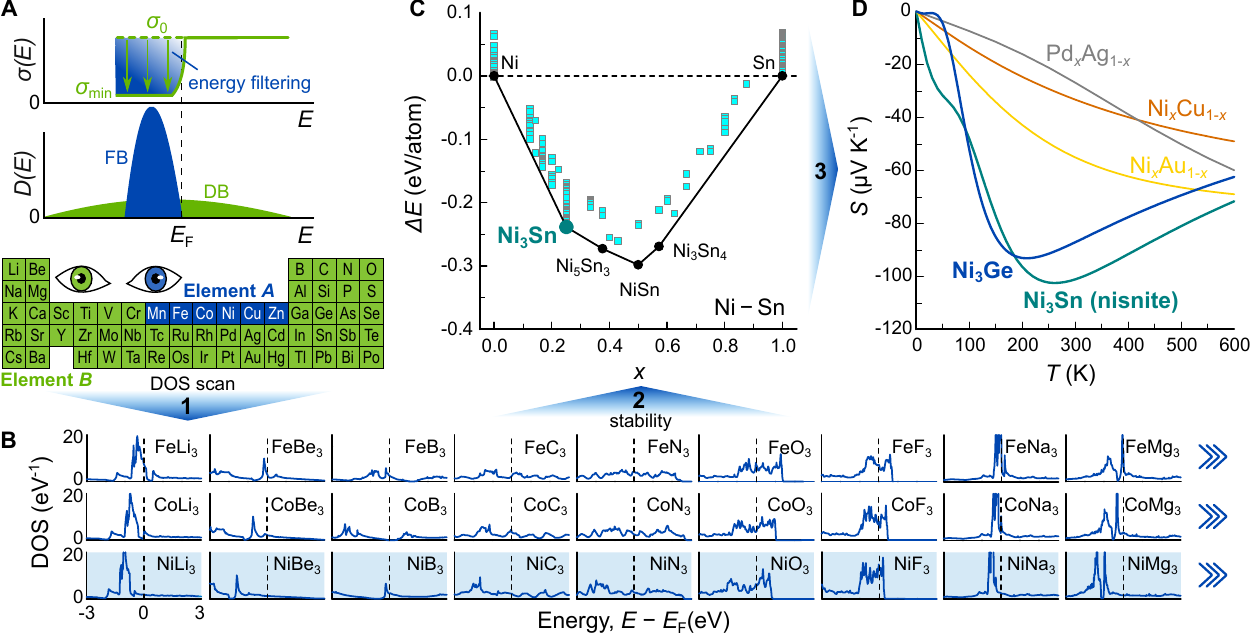}
	\caption{\textbf{Density of states screening for novel high-performance metallic thermoelectrics enabled by intrinsic energy filtering.} (\textbf{A}) Schematic of IEF in gapless systems hosting a combination of flat (blue) and dispersive (green) bands at the Fermi level $E_\text{F}$. In step 1, a broad range of 255 fictitious binary cubic systems was scanned with a fixed stoichiometry ($A_1B_3$), where $A$ is a transition metal from Mn to Zn, providing localized $3d$ states and $B$ is any element from Li to La. (\textbf{B}) DFT-calculated energy-dependent DOS near $E_\text{F}$ for selected $A_1B_3$ binaries from step 1. In step 2, a more detailed screening of Ni binaries was performed with variable-composition crystal structure prediction calculations, taking into account thermodynamic stability. (\textbf{C}) shows the formation energy as a function of atomic fraction for the binary Ni-Sn system identified in step 2. In a last step, the temperature-dependent Seebeck coefficient $S(T)$ was calculated, using an interband scattering model, which estimates $\tau^{-1}(E) \propto D(E)$ as per Fermi's golden rule. (\textbf{D}) Seebeck coefficient of two new systems, L1$_2$-ordered \ce{Ni3Ge} and \ce{Ni3Sn}, identified from our multi-step screening, compared to group 10 and 11 transition metal alloys ($x=0.5$).} 
	\label{Fig2}
\end{figure*}

Here, we employ intrinsic energy filtering (IEF) as a new strategy for achieving high $P\!F$ driven by selective carrier scattering in flat-band (FB) materials (Fig.\,\ref{Fig1}). More specifically, when flat and dispersive bands (DBs) are engineered to the Fermi level $E_\text{F}$, charge carriers from the DB can transition into the FB in the overlapping energy range by scattering off impurities, phonons or even other electrons. This enables a large Seebeck effect even in metals with a high carrier concentration that would otherwise display a negligibly small $S$. The effectiveness of IEF is largely determined by the density of states (DOS), which gives the total available phase space for possible interband transitions and by the respective scattering potentials \cite{PRXEnergy.3.043009}.

%\vspace*{-0.3cm}
%\section*{Screening strategy}
%\vspace*{-0.1cm}
%In a first step, the DOS of all solid elements was gathered from the Materials Project database \cite{jain2013commentary} and summarized in a periodic table to get a good overview (Fig.\,1A). For an initial rough screening, we calculated the DOS of fictitious binary systems in a cubic crystal structure and fixed stoichiometry (Methods), without taking into account material stability or experimentally confirmed existing phases. We reckon that this leaves a lot of room for improvement, e.g., by making use of other databases, such as the Inorganic Crystal Structure Data Base (ICSD) \cite{bergerhoff1983inorganic} or the recently released GNoMe database by the Google Deepmind team \cite{merchant2023scaling}. Transition metals ranging from Mn to Zn were chosen as starting elements providing the rather localized 3$d$ states (flatter bands), while the second element could be any element ranging from Li to La (Fig.\,\ref{Fig1}b).

\vspace*{-0.3cm}
\section*{RESULTS}
\vspace*{-0.1cm}
\noindent We developed a multi-step screening method, solely based on the DOS calculated by density
functional theory (DFT), to search for materials with the potential for IEF (Fig.\,\ref{Fig2}A). Our goal is to maximize IEF by engineering non-dispersive bands (very sharp DOS) right below/above $E_\text{F}$ in a metallic background of dispersive conduction bands. This requires a metallic material with non-hybridizing, localized states (e.g. 3$d$ orbitals). If this is the case, the energy of such localized $d$ orbitals depends mainly on the relative chemical potential of the two atomic species (weighted by composition), and only minorly on the details of the crystal structure.

Following this idea, we start in step 1 with a rough screening of fictitious binary systems
using a simple cubic unit cell and $A_1B_3$ composition (Methods). The DOSs of all possible
compounds with $A$ being a transition metal ranging from Mn to Zn and $B$ being any element from Li to La were calculated by DFT. As shown in Fig.\,\ref{Fig2}B, the sharp peaks associated with localized $3d$ states are mostly unperturbed, except for a shift in energy. Initial screening highlights binary systems with Fe, Co, or Ni as the transition metal providing the non-dispersive bands and $B$ elements from the alkali metals, alkaline earths, or the III$^\text{rd}$ and IV$^\text{th}$ main groups as promising. At this point, it is worth mentioning that the search for flat-band hosting materials is a highly active field of research, with several alternative strategies being explored beyond the use of localized $3d$ (or $f$) orbitals. For example, destructive phase interference in frustrated lattices \cite{kang2020topological} and the formation of moiré superlattices, such as in twisted bilayer graphene \cite{xie2021fractional}, have emerged as promising routes and even dedicated flat-band databases have been developed \cite{regnault2022catalogue,neves2024crystal}. Thus, we reckon that our screening method has potential for future improvement and expansion by making use of these newly developed flat-band catalogues.

\begin{figure*}[tbh]
\newcommand{\setwidth}{0.45}
			\centering
			\hspace*{0cm}
		\includegraphics[width=0.9\textwidth]{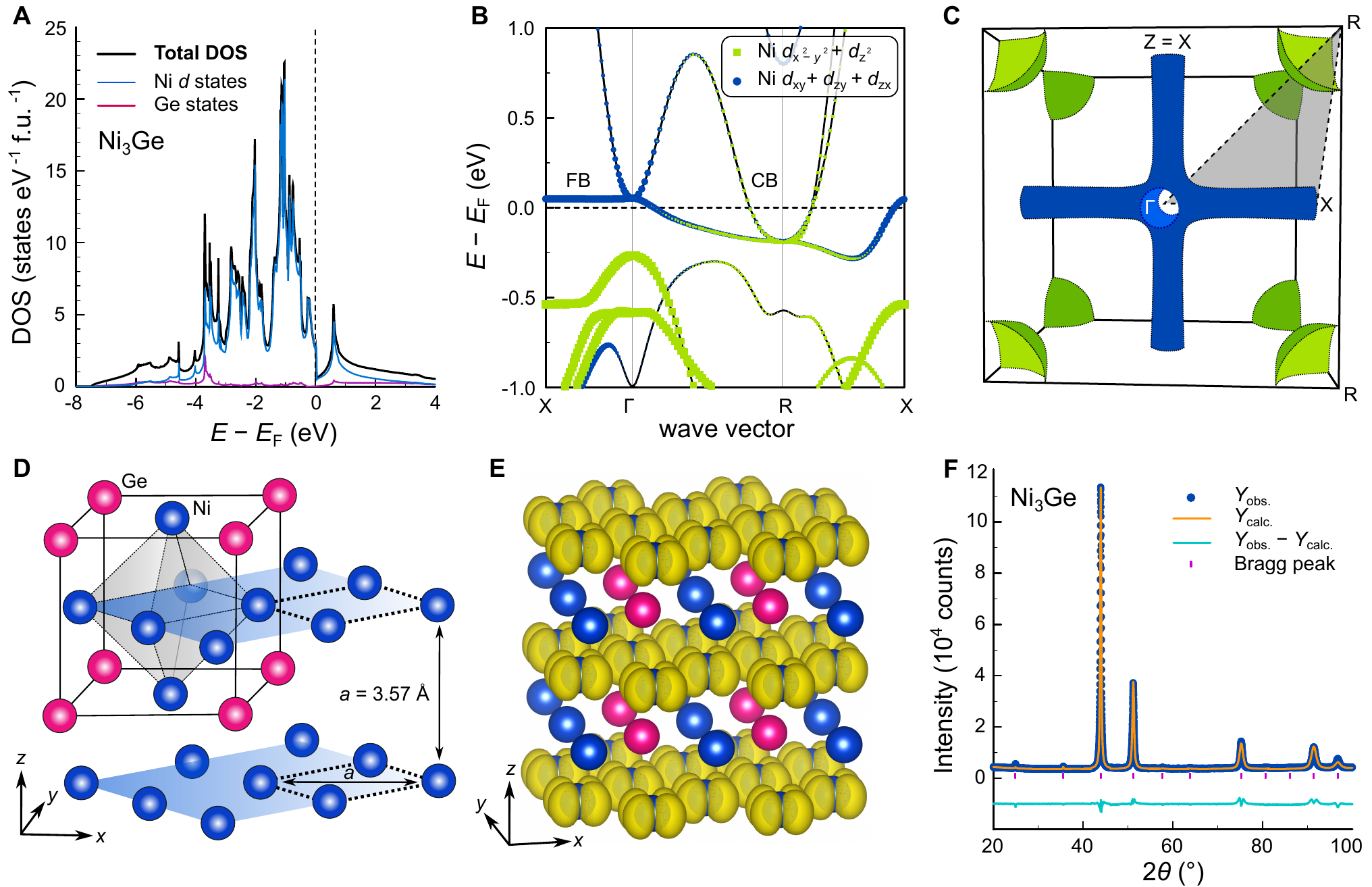}
	\caption{\textbf{Electronic structure and atomic-like flat bands in L1$_2$-ordered \ce{Ni3Ge}.} (\textbf{A}) Total and partial DOS of \ce{Ni3Ge}, showing dominant Ni $3d$ states near %the Fermi energy 
	$E_\text{F}$. (\textbf{B}) Band structure of \ce{Ni3Ge} with orbital-decomposed Ni $3d$ states, revealing a flat band along the $\Gamma$\,--\,X direction, just 20\,meV above $E_\text{F}$. (\textbf{C}) Fermi surface of \ce{Ni3Ge}, showing low-dimensional tubular features along $\Gamma$\,--\,X and a doubly degenerate pocket at the R point, corresponding to partially filled conduction bands. (\textbf{D}) Crystal structure of \ce{Ni3Ge}, crystallizing in the ordered \ce{Cu3Au}-type structure. (\textbf{E}) Charge density $n(r)$ for the flat band at X, illustrating localized $d_{xy}$ orbitals within the $xy$ planes. (\textbf{F}) X-ray diffraction pattern of \ce{Ni3Ge}, with Rietveld refinement confirming successful synthesis of single-phase L1$_2$-ordered \ce{Ni3Ge}.} 
	\label{Fig3}
\end{figure*}

The second step of our screening consisted of a series of variable-composition crystal structure prediction calculations
(Methods) for the most promising pairs of elements emerging from the initial screening in step 1. We focused on Co- and Ni-based binaries with III$^\text{rd}$ and IV$^\text{th}$ main group elements, which are experimentally more convenient than alkalis and alkaline earths and exhibit a plethora of stable phases. Figure\,\ref{Fig2}C shows the calculated formation energies of the binary Ni-Sn system. Among the over forty thermodynamically stable compounds we found, particularly promising is the family of $A_3B$ compounds, crystallizing in the L1$_2$-ordered \ce{Cu3Au} structure, where $A$ is a Ni-group element and $B$ is Ge, Sn, or Pb. These gapless materials exhibit intriguing low-dimensional electronic structures, with remarkably flat, atomic-like bands in certain Brillouin zone directions, yet they have entirely flown under the radar of the TE community, which has thus far heavily focused on narrow-gap semiconducting systems. 

In a third step, we employed a simple interband scattering model, wherein the carrier scattering rate is estimated to be proportional to the density of states (as per Fermi's golden rule) $\tau^{-1}(E)\propto D(E)$, to calculate the temperature-dependent Seebeck coefficient $S(T)$ of our newly discovered candidate materials. The same model has previously been shown to work well for similar systems, such as binary group 10 and 11 transition metal alloys \cite{garmroudi2023high} and the chiral semimetal CoSi \cite{xia2019high}. Figure\,\ref{Fig2}D displays $S(T)$ of $L1_2$-ordered \ce{Ni3Ge} and \ce{Ni3Sn} estimated in this manner. The latter is in fact a natural mineral ("nisnite"), which has been discovered in Canadian mines, where it is believed to have formed under elevated pressure and temperature conditions (2.5\,--\,4.5\,kbar and 563\,--\,673\,K) in Earth's crust \cite{rowe2011nisnite}. While nisnite would be the most promising candidate, it can only be synthesized under high pressure. On the other hand, \ce{Ni3Ge} is stable at ambient pressure and across the entire temperature range up to its melting point at around $1400\,$K, and also cheaper than Pd- and Pt-containing systems, whose electronic structures and estimated $S(T)$ curves are shown in fig.\,S9. All these factors motivated us to perform an in-depth experimental and theoretical study of \ce{Ni3Ge}.

%\vspace*{-0.3cm}
%\section*{Electronic structure of \ce{Ni3Ge}}
%\vspace*{-0.1cm}
\noindent Figure\,\ref{Fig3} summarizes the distinguished electronic structure features of \ce{Ni3Ge}. In Fig.\,\ref{Fig3}A it can be seen that the majority of the DOS can be attributed to the $3d$ states of Ni. Most prominently, around $E_\text{F}$ there are two dispersive conduction bands: one slightly above $E_\text{F}$ at $\Gamma$ and another partially filled one at the R point (Fig.\,\ref{Fig3}B). Additionally, an exceptionally flat band (FB) occurs along $\Gamma$\,--\,X, which remains rather flattened along other $k$ directions as well. This yields peculiar low-dimensional features (tubes) in the Fermi surface (Fig.\,\ref{Fig3}C), which are accompanied by the doubly degenerate pocket at R. We note that both the Fermi surface tubes and the FB along $\Gamma$\,--\,X, from which they originate, display striking similarity to the electronic structures found in some full-Heusler compounds, where very large power factors were previously predicted theoretically \cite{bilc2015low}.
\begin{figure*}[tbh]
\newcommand{\setwidth}{0.45}
			\centering
			\hspace*{0cm}
			\includegraphics[width=0.95\textwidth]{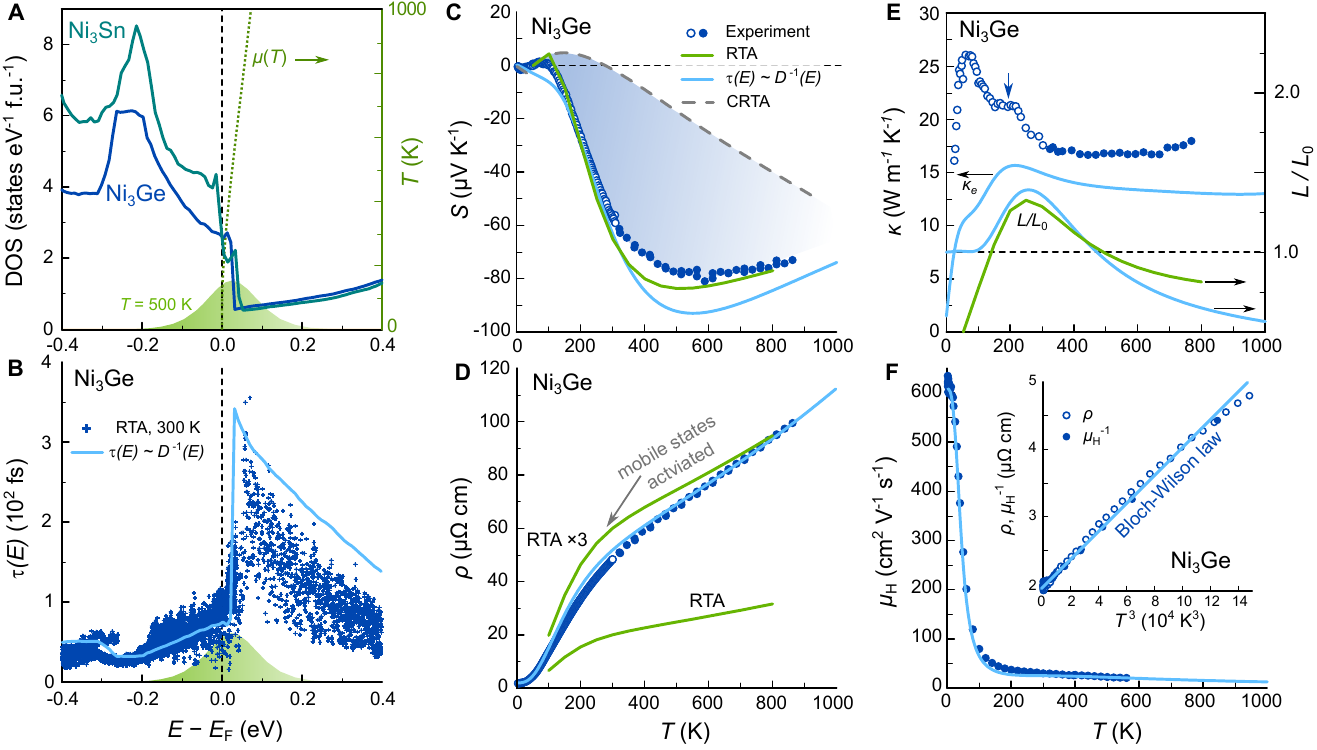}
	\caption{\textbf{Flat-band induced conductivity edge and electronic transport signatures of energy filtering in \ce{Ni3Ge}.} (\textbf{A}) DOS of \ce{Ni3Ge} and \ce{Ni3Sn} as calculated by DFT, showing a sharp step near $E_\text{F}$. The dotted line marks the temperature-dependent chemical potential $\mu(T)$, and the green area represents the derivative of the Fermi-Dirac distribution $(-df/dE)$ at 500\,K, about where $|S|$ reaches its maximum. (\textbf{B}) Energy-dependent electron lifetimes, obtained from \textit{ab initio} calculations of electron-phonon interaction and scattering matrix elements in the relaxation time approximation (RTA) and using a simple $D^{-1}(E)$ model. (\textbf{C} to \textbf{F}) Temperature-dependent Seebeck coefficient $S(T)$, electrical resistivity $\rho(T)$, thermal conductivity $\kappa(T)$ and Hall mobility $\mu_\text{H}(T)$ of \ce{Ni3Ge} with theoretical calculations in the constant relaxation time approximation (green dashed line), the RTA (green solid line) and the $D^{-1}(E)$ model (blue solid line). Blue shaded area in \textbf{D} refers to the contribution from IEF. Inset in \textbf{F} shows that the scattering rate follows a cubic behavior at low temperatures, as expected for electron--phonon interband scattering.} 
	\label{Fig4}
\end{figure*}
Like for the Heusler compounds, it can be shown that these low-dimensional electronic structure features originate from a lack of $dd$ hopping in certain crystal directions. Figure\,\ref{Fig3}D shows the highly ordered \ce{Cu3Au}-type crystal structure of \ce{Ni3Ge}, where Ni atoms form an octahedron inside a cube with Ge atoms at the corners. An extended version shows that there are Ni layers separated by a lattice parameter of $a=0.357\,$nm, comparable to that of elemental \textit{fcc} Ni ($a=0.352\,$nm). In Fig.\,\ref{Fig3}E, we plot the charge density $n(r)$ of the FB at X, as obtained from our DFT calculations. The shape of $n(r)$ resembles atomic-like $d_{xy}$ orbitals, signifying that there is no $dd$ hopping across the layers and that the bonding for these electronic states is two-dimensional. The same holds true, \textit{mutatis mutandis}, for the $yz$ and $xz$ planes. We successfully synthesized phase-pure \ce{Ni3Ge} in the \ce{Cu3Au} structure (Fig.\,\ref{Fig3}F) and studied the TE properties in a broad temperature range (2\,--\,860\,K) as discussed next. 

%\vspace*{-0.3cm}
%\section*{Signatures of intrinsic energy filtering}
%\vspace*{-0.1cm}
\noindent In Fig.\,\ref{Fig4}, we demonstrate how IEF occurs from interband scattering in \ce{Ni3Ge} and manifests itself in all of the temperature-dependent transport properties. First, the rapid variation and steep edge in the DOS associated with the atomic-like FB, is depicted in Fig.\,\ref{Fig4}A (DOS of \ce{Ni3Sn} is shown for comparison as well). The temperature-dependent chemical potential $\mu(T)$ and the derivative of the Fermi-Dirac distribution $(-df/dE)$ at 500\,K, about where $S$ displays its maximum values, are plotted over the DOS, highlighting that the entire relevant energy range can be readily excited at temperatures accessible by our transport measurements. The steep edge in the DOS implies that also the scattering phase space varies rapidly as a function of energy. Figure\,\ref{Fig4}B compares the energy-dependent carrier relaxation time, estimated solely from the DOS via $\tau(E)\propto D^{-1}(E)$, %versus the results obtained from much more expensive and 
with advanced \textit{ab initio} calculations of electron-phonon interactions and scattering rates in the relaxation time approximation (RTA)\cite{cepellotti2022phoebe}; details in Methods. It is evident that the simple $D^{-1}(E)$ model captures the energy-dependent behavior of $\tau(E)$ well, reproducing its sharp peak just above $E_\text{F}$. Here, the dispersive conduction band at R and the flat bands overlap (Fig.\,\ref{Fig3}B) and $\tau(E)$ drops rapidly due to interband scattering. 

Notably, in pristine \ce{Ni3Ge}, interband scattering is intrinsically mediated via phonons (as sketched in Fig.\,\ref{Fig1}) and possibly electron-electron correlations, which differs markedly from the mechanism in binary \ce{Ni_xCu_{1-x}} and \ce{Ni_xAu_{1-x}} alloys, where extrinsic impurity/disorder scattering occurs between $s$-like conduction electrons and randomly distributed Ni atoms in the alloy (fig.\,S11) \cite{garmroudi2023high}. This has some implications, since phonons can transfer their momentum onto the charge carriers. When the phonon wave vector $q$ is small (at low temperatures), extensive interband transitions between carrier pockets at different points in the Brillouin zone is severely limited, as they are far apart in reciprocal space. This is indeed apparent in the low-temperature behavior of $S(T)$, which starts off with a small positive slope that is not captured in the $D^{-1}(E)$ model. Once $T$ reaches about one third of the Debye temperature $\Theta_\text{D}\approx 460(20)\,$K, high-$q$ phonons can scatter low-energy charge carriers from the R point into the FB, yielding a sign reversal and a strong enhancement of $|S(T)|$. Experimental $S(T)$ data can be excellently reproduced by our RTA calculations in the entire temperature range because they incorporate electron-phonon scattering, whereas the $D^{-1}(E)$ model gives very good agreement only above $\Theta_\text{D}/3$ where all phonon modes are accessible. On the contrary, when $S(T)$ is calculated in the commonly utilized constant relaxation time approximation (CRTA), where $\tau(E)$ is treated as a constant that drops out in the Boltzmann transport expression for $S(T)$, the experimental trend is not reproduced and absolute values are severely underestimated. This highlights the pivotal role of IEF by interband scattering processes in enabling a large Seebeck effect in metallic \ce{Ni3Ge}. The blue shaded area in Fig\,\ref{Fig4}C corresponds to the contribution of IEF, which yields a $\approx 300$\,\% enhancement of $|S|$ at the temperature where it reaches its maximum. The deviation from CRTA becomes even greater at 300\,K, where the Seebeck effect is entirely attributed to IEF.

Figure\,\ref{Fig4}D shows the temperature-dependent electrical resistivity $\rho(T)$ of \ce{Ni3Ge}. Interestingly, the RTA framework qualitatively reproduces the temperature dependence of $\rho(T)$, but underestimates the absolute values by about a factor of three. Since our samples are polycrystalline, this discrepancy could be related to grain boundary scattering, although this seems unlikely given the quite large residual resistivity ratio $\rho_\text{300\,K}/\rho_\text{4\,K} \approx 26$. Another possibility could be that the FB at $E_\text{F}$ leads to a renormalization of the electron-phonon coupling and/or the quasiparticle effective mass, which is not captured by our DFT-based calculations. Such effects would be much more relevant for the resistivity since global enhancement factors of the scattering rate drop out in the Boltzmann transport expression for $S(T)$ \cite{cepellotti2022phoebe}.

%To further demonstrate the validity of the simple $D^{-1}(E)$ model, we calculated both the electrical resistivity $\rho(T)$ as well as the Hall mobility $\mu_\text{H}(T)$ by fitting the energy-independent term, that drops out for $S(T)$ but does not for $\rho(T)$ and $\mu_\text{H}(T)$. The energy-independent term of the scattering rate can be described by the well-known Bloch-Gr\"uneisen law 
%
%\begin{equation}
%\label{BG}
%\tau^{-1}_\text{e-p}(T) \sim  \mathcal{A} \left(\frac{T}{\Theta_\text{D}}\right)^n\,\int_0^{\Theta_\text{D}/T}\frac{x^n}{\left(e^x-1\right)\left(1-e^{-x}\right)}dx\,.
%\end{equation} 
%
%\noindent Here, $\mathcal{A}$ is a constant that depends -- apart from natural constants -- on the electron-phonon coupling, the DOS around $E_\text{F}$ and the carrier velocities. As the Debye temperature is known from specific heat measurements (see SI), only $\mathcal{A}$ has to be fitted. The results of these single-parameter fits are shown as dashed lines in Fig.\,\ref{Fig3}E and F, revealing outstanding agreement with experimental data. This fully confirms that the energy-dependent behavior of $\tau(E)$ is accurately represented by assuming $\tau(E)\propto D^{-1}(E)$. 

Additional proof for the importance of phonon-mediated interband scattering is seen in the low-temperature behavior of $\rho(T)$, the Hall mobility $\mu_\text{H}(T)$ (inset Fig.\,\ref{Fig4}F) and even in the thermal conductivity $\kappa(T)$ (Fig.\,\ref{Fig4}E), which shows a shoulder-like feature around 200\,K (blue vertical arrow) in agreement with theoretical calculations predicting a peak in the temperature-dependent Lorenz number owing to thermal activation of carriers across the scattering-induced conductivity edge. 

As described by the well-known Bloch-Gr\"uneisen law, the scattering rate for electron-phonon scattering becomes linear at high temperatures, and at low temperatures $\tau^{-1}(T)$ follows a power law $T^n$, where $n=5$ for common metals and $n=3$ in the case of strong interband scattering (Bloch-Wilson limit) \cite{wilson1938electrical}. Indeed, $\rho(T)$ and $\mu_\text{H}^{-1} (T)$, both of which are reflecting the scattering rate as the carrier concentration is almost temperature-independent (fig.\,S10), follow a cubic power law at low temperatures (inset Fig.\,\ref{Fig4}F).

\begin{figure*}[t!]
\newcommand{\setwidth}{0.45}
			\centering
			\hspace*{0cm}
		\includegraphics[width=1\textwidth]{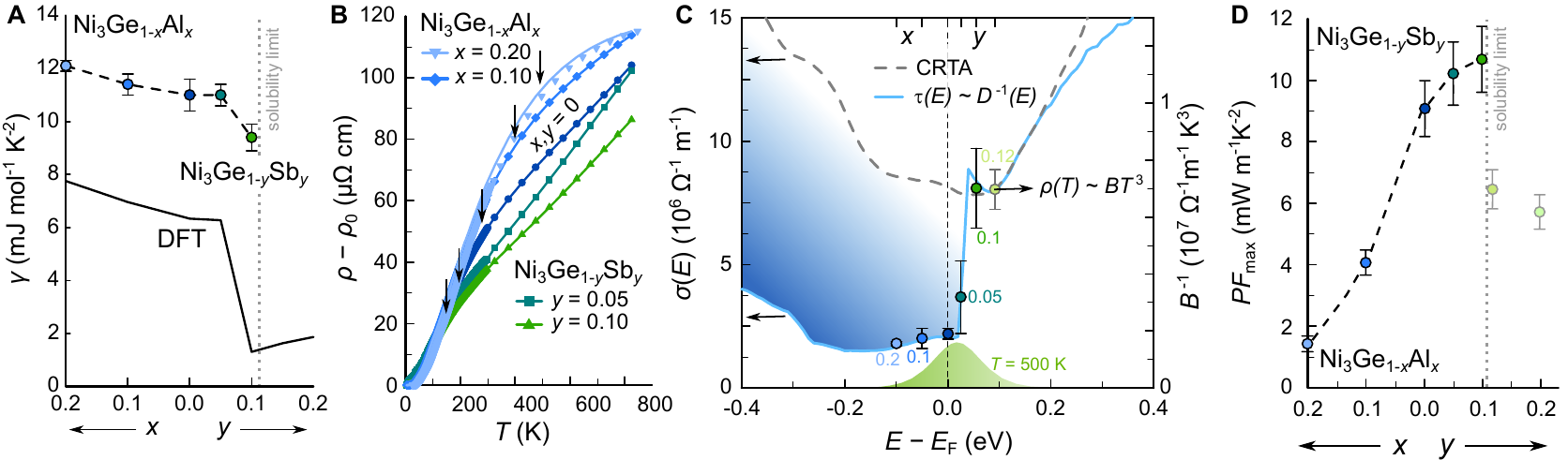}
	\caption{\textbf{Tuning the Fermi level with respect to the flat-band states of \ce{Ni3Ge}.} (\textbf{A}) Sommerfeld coefficient $\gamma$, extracted from the specific heat $C/T$ at $T\rightarrow 0$ of $p$-doped \ce{Ni3Ge_{1-x}Al_x} and $n$-doped \ce{Ni3Ge_{1-x}Sb_x}, compared with DFT calculations assuming rigid-band-like doping. (\textbf{B}) Temperature-dependent electrical resistivity of \ce{Ni3Ge_{1-x}Al_x} and \ce{Ni3Ge_{1-x}Sb_x}, showing a consistent decrease in resistivity as the nominal valence electron concentration increases, pushing $E_\text{F}$ into more dispersive bands and away from the flat band. Moreover, a shoulder-like feature in $\rho(T)$, corresponding to thermal activation of dispersive-band charge carriers across the conductivity edge, appears and is shifted towards higher/lower temperatures with Al/Sb doping (arrows in Fig.\,\ref{Fig5}B). (\textbf{C}) Energy-dependent transport distribution function from DFT, assuming a constant relaxation time (CRTA; dashed line) and $\tau(E) \propto D^{-1}(E)$ (solid line), with interband scattering filtering out low-energy carriers (blue shaded area). The step-like feature in $\sigma(E)$ can be experimentally reconstructed by determining $\sigma(E_\text{F})\propto B^{-1}$ from the cubic power law $\rho(T)\sim BT^3$ at low temperatures. (\textbf{D}) Compostition-dependent maximum power factor of \ce{Ni3Ge_{1-x}Al_x} and \ce{Ni3Ge_{1-y}Sb_y}.} 
	\label{Fig5}
\end{figure*}

%\vspace*{-0.3cm}
%\section*{Tunability of Fermi level}
%\vspace*{-0.1cm}
\noindent To tune the position of $E_\text{F}$ with respect to the atomic-like flat band, we investigated the effects of chemical doping in \ce{Ni3Ge} (Fig.\,\ref{Fig5}). Similar to semiconductors, tuning the position of $E_\text{F}$ relative to the scattering-induced conductivity edge is crucial to assess optimal performance in scattering-tuned metals. Myriad materials crystallize in the \ce{Cu3Au} structure, with 1,247 entries in the Inorganic Crystal Structure Data Base as of November 2024, allowing for outstanding tunability and flexibility, e.g., when it comes to adjusting $E_\text{F}$ through aliovalent alloying. For example, Al substitution at the Ge site in \ce{Ni3Ge_{1-x}Al_x} lowers $E_\text{F}$, shifting it into the higher DOS, while Sb doping in \ce{Ni3Ge_{1-x}Sb_x} raises $E_\text{F}$ above the edge into the dispersive bands. Figure\,\ref{Fig5}A shows that the Sommerfeld coefficient $\gamma$, extracted from low-temperature specific heat measurements (fig.\,S12A), follows a trend consistent with the DOS calculated by DFT, assuming a rigid-band shift of $E_\text{F}$, which confirms the effectiveness of the doping. We find, however, that the experimental $\gamma$ values are enhanced compared to DFT predictions, even when accounting for electron-phonon coupling calculated by Density Functional Perturbation Theory (fig.\,S7C), yielding $\lambda_\text{DFT} < 0.1$ versus  $\lambda_\text{exp} \approx 0.74(9)$. This suggests a renormalization of either $\lambda_\text{e-p}$ or the effective mass by many-body effects, affecting $\rho(T)$ and $\gamma$, but not $S(T)$.

Figure\,\ref{Fig5}B displays $\rho(T)-\rho_0$, for \ce{Ni3Ge_{1-x}Al_x} and \ce{Ni3Ge_{1-x}Sb_x}. Electron doping (Ge/Sb) reduces the slope of $\rho(T)$, while hole doping (Ge/Al) steepens it, reflecting shifts of $E_\text{F}$ into the dispersive and flat bands, respectively. At higher temperatures, the slopes of $\rho(T)$ decrease, associated with thermal activation of mobile carriers above the conductivity edge. This shoulder, indicated by vertical black arrows in Fig.\,\ref{Fig5}B, shifts to higher temperatures with Al doping (as $E_\text{F}$ moves away from the conductivity edge) and to lower temperatures with Sb doping (as $E_\text{F}$ approaches the conductivity edge).
We are even able to experimentally reconstruct the scattering-induced conductivity edge in the energy-dependent transport distribution function $\sigma(E)$ by carefully analyzing $\rho(T)$ of the doped samples (Fig.\,\ref{Fig5}C). At low temperatures, the aforementioned Bloch-Wilson limit yields $\rho(T)\sim BT^3$, where $B$ is determined by the carrier velocities and DOS, encapsulated in $\sigma(E_\text{F})$ and by other physical parameters, such as the electron-phonon coupling constant $\lambda_\text{e-p}$. Assuming the latter do not change dramatically with doping, $\sigma(E_\text{F})$ can be extracted from the cubic fits of our experimental data, in excellent agreement with the theoretical estimate of $\sigma(E)$.

The temperature-dependent power factor, Seebeck coefficient, thermal conductivitiy and $zT$ of \ce{Ni3Ge_{1-x}Al_x} and \ce{Ni3Ge_{1-x}Sb_x} are shown in figs.\,S13A to D. $S(T)$ decreases with Al doping as electron-hole asymmetry diminishes (cf. Fig.\,\ref{Fig5}C), while Sb doping causes a slight shift of the maximum Seebeck coefficient to lower temperatures. The maximum $P\!F$, exceeding 10\,\PFunit, is achieved in \ce{Ni3Ge_{0.9}Sb_{0.1}} near the Sb solubility limit at 300\,--\,400\,K, with a peak $zT \approx 0.3$; note that the majority of waste heat streams arise in this temperature range \cite{forman2016estimating}. As shown in fig.\,S13E, these values are -- apart from the recently discovered Ni-Au alloys \cite{garmroudi2023high} -- substantially larger than those found in the few known systems \cite{li2005effects,kanazawa2012band,mao2015high}, where IEF enhances TE performance (comparison to semiconductors in fig.\,S14).

\vspace*{-0.3cm}
\section*{DISCUSSION}
\vspace*{-0.1cm}
\noindent In conclusion, we developed a multi-step DFT-based screening method for identifying novel gapless TEs with the potential for intrinsic energy filtering. When flat and dispersive bands overlap, charge carriers can be selectively immobilized through interband scattering. This process is largely dictated by the DOS, which determines the total available scattering phase space. Taking the DOS as a simple descriptor for IEF, we discovered a new family of binary metallic compounds, crystallizing in the \ce{Cu3Au}-type structure, with ultrahigh $P\!F>10\,$\PFunit\, in \ce{Ni3Ge_{1-x}Sb_x} around room temperature ($PF_\text{max}\approx 11\,$mW\,m$^{-1}$\,K$^{-2}$ at 375\,K). 

In the future, an important question to address will be how to further pile up and sharpen the DOS to enhance the phase space for interband scattering, e.g. by further flattening the bands or via band convergence. Different from semiconducting systems, however, $\sigma(E)$ should not be locally enhanced but reduced by the converged bands. Graziosi et al. recently derived a set of design criteria from a two-parabolic band model, showing that, aside from the DOS effective mass asymmetry, the respective deformation potentials and the band overlap are also crucial parameters to maximize the effectiveness of interband scattering for metallic thermoelectrics \cite{PRXEnergy.3.043009}.

\begin{figure}[t!]
\newcommand{\setwidth}{0.45}
			\centering
			\hspace*{0cm}
		\includegraphics[width=0.35\textwidth]{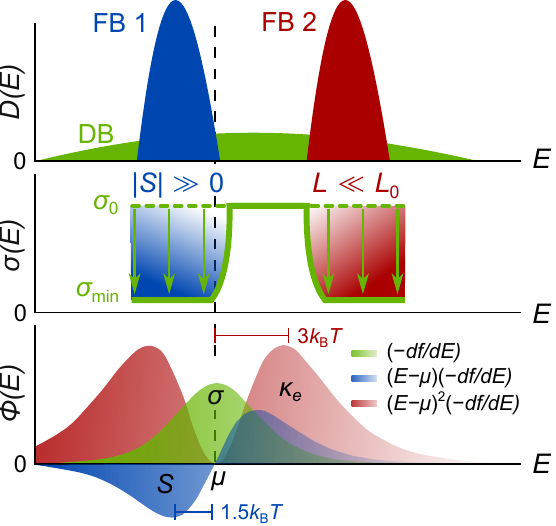}
	\caption{\textbf{Towards the optimal transport function via interband scattering.} Filtering low-energy charge carriers with a FB just below $E_\text{F}$ (1) enables a large Seebeck effect while filtering high-energy charge carriers further above $E_\text{F}$ with another FB (2) restricts electronic heat transport without affecting the electrical conductivity, which reduces the Lorenz number and enhances $zT$.} 
	\label{Fig6}
\end{figure}

Furthermore, an interesting direction to explore would be to selectively filter out high energy charge carriers further above $E_\text{F}$ to create a boxcar-shaped $\sigma(E)$ (Fig.\,\ref{Fig6}). As suggested by recent theoretical studies \cite{park2021optimal,ding2023best}, this would be the optimal transport function, superior to the delta function proposed originally by Mahan and Sofo in their seminal work \cite{mahan1996best}. Here, we demonstrated that immobilizing low-energy carriers just below $E_\text{F}$ is crucial for enhancing $S$. Engineering flat bands further above $E_\text{F}$, on the other hand, can filter out charge carriers that most strongly contribute to thermal transport, while retaining a high $\sigma$ \cite{mckinney2017search}. This way, the Lorenz number $L$ could be reduced well below the Sommerfeld value $L_0=(\pi^2/3)(k_\text{B}/e)^2$, which would further enhance $zT$, particularly in these gapless systems as $\kappa_e \gg \kappa_l$. 

With the development of increasingly advanced materials databases \cite{regnault2022catalogue}, the advent of machine-learning-assisted material discovery \cite{merchant2023scaling}, and significant progress in theoretical transport calculations beyond the CRTA \cite{ganose2021efficient,cepellotti2022phoebe,graziosi2023electra}, the search for high-performance TEs with IEF has become more viable and accessible than ever. Our work demonstrates that a simple, easily computed property like the density of states can be a powerful tool for identifying promising candidate materials, and confirms that TE materials with ultrahigh $P\!F=10^1$\,--\,10$^2\,$\PFunit, long thought to be rare exceptions, can be readily designed and may be far more common than initially expected when exploring gapless metallic systems instead of the typically studied narrow-gap semiconductors.
\vspace*{-0.3cm}
\section*{MATERIALS AND METHODS}
\vspace*{-0.1cm}
\section*{Synthesis of starting materials}
\vspace*{-0.2cm}
\noindent High-purity bulk elements (Ni 99.99\,\%, Ge 99.999\,\%, Al 99.999\,\%, Sb 99.999\,\%) were carefully weighed in stoichiometric amounts, and polycrystalline ingots were produced using high-frequency induction melting in a water-cooled copper coldboat under an argon atmosphere. It is worth mentioning that although Sb is typically prone to evaporation due to its low vapor pressure, this was not a concern for the materials used in this process and the mass loss was negligibly small for all samples ($\delta m/m < 0.001$). Although, \ce{Ni3Ge}-based compounds melt almost congruently into a single phase, small impurities were observed directly after the melt synthesis. After annealing the samples at 973\,K for several days, the samples were homogenized and phase-pure specimen were obtained, which also showed a notably higher Seebeck coefficient, than those measured directly after the melt synthesis. The sample ingots were cut using a high-speed cutting machine (Accutom from Struers). Rectangular samples with a typical geometry of $10\times 1.5\times 1.5$\,mm were obtained and offcuts of the samples were crushed and ground by hand to a fine powder to be used for powder x-ray diffraction measurements.
\vspace*{-0.3cm}
\section*{Structural characterization}
\vspace*{-0.2cm}
\noindent X-ray powder diffraction measurements were conducted at the Institute of Solid State Physics, TU Wien, using an in-house diffractometer (AERIS by PANalytical). These measurements utilized standard Cu K-$\alpha$ radiation, with data collected in the Bragg-Brentano geometry over the angular range $20^\circ < 2\theta < 100^\circ$. Rietveld refinements on the obtained powder patterns were performed using the program PowderCell.
\vspace*{-0.3cm}
\section*{High-temperature transport measurements} 
\vspace*{-0.2cm}
\noindent The bar-shaped samples were mounted in a commercial setup (ZEM3 by ULVAC-RIKO) and the electrical resistivity and Seebeck coefficient were measured as a function of temperature. The setup operates from room temperature up to $860\,$K (or higher) and utilizes an infrared furnace. The sample chamber is evacuated and some He exchange gas ensures thermal coupling to the sample. Fig.\,S1A showcases the measurement principle and setup, where the sample is placed and mechanically fixed between two Pt electrodes. A temperature gradient is generated across the sample by a heater at the bottom. The temperature difference $\Delta T$ and voltage $\Delta V$ are measured by two thermocouples TC1 and TC2 with an approximate distance $d=3$\,mm between TC1 and TC2. Since spurious voltages can occur, the measurement is typically conducted $N$ times at $N$ different values of $\Delta T$, and a linear regression is performed to determine the Seebeck coefficient, which -- after subtraction of $S_\text{wire}$, that is, the Seebeck coefficient of the thermocouple -- comes out as the slope of the linear fit

%\begin{equation}
%S = -\frac{\left(\sum_{i=1}^N \Delta T_i\right)\left(\sum_{i=1}^N \Delta V_i\right)-N\left(\sum_{i=1}^N \Delta T_i \Delta V_i\right)}{\left(\sum_{i=1}^N \Delta T_i\right)^2-N\left(\sum_{i=1}^N \Delta T_i^2\right)}\,.
%\end{equation}

Usually, $N=3$ and $\Delta T = 10,\,15$ and 20\,K is assumed to reach a desired threshold of accuracy. Note that this represents the temperature difference with respect to the bottom heater. The real temperature difference measured at the sample is usually an order of magnitude smaller. Additionally the resistivity can be determined simultaneously at each temperature by making use of a four-probe technique.

Temperature-dependent measurements of the thermal conductivity $\kappa(T)$ in the temperature range 300\,K up to approximately $760\,$K have been conducted by making use of a commercially available laser/light flash setup (LFA 500 by Linseis). Pictures of the setup and sample holder, as well as a schematic of the measurement principle are shown in fig.\,S2. The thermal conductivity is given by $\kappa=D\,\rho_\text{m}\,C$, where $D$ denotes the thermal diffusivity, $\rho_\text{m}$ the material density and $C$ the specific heat. The thermal diffusivity and the specific heat can be simultaneously measured by the device, while the material density exists as an input parameter that can be determined via Archimedes' principle. The specific heat is measured via a differential scanning calorimetry method by making use of an appropriate reference sample. Fig.\,S2C shows a sketch of the measurement principle for the thermal diffusivity. In the LFA 500, a xenon flash lamp heats the sample from the sample bottom. The absorbed heat is conducted through the sample and reaches the top surface where the temperature rises and heat is radiated from a small hole. The signal is then picked up by a detector further away from the sample and converted to a voltage, which is recorded as a function of time. Fig.\,S2C depicts a typical measurement curve, where the signal intially spikes due to ballistic heat transport, after the light/laser flash is fired, followed by a steady increase up to a maximum value. The time to the half maximum $t_{1/2}$ is directly related to the thermal diffusivity $D$ via $t_{1/2} \propto d^2/D$, where $d$ is the thickness of the sample (usually $1-2$\,mm). 

\vspace*{-0.3cm}
\section*{Low-temperature transport measurements}
\vspace*{-0.2cm}
\noindent The thermoelectric characterization at low temperatures was carried out at on the same rectangular bar-shaped sample pieces that were used for the high-temperature measurements. The temperature-dependent electrical resistivity was measured using a custom-built sample probe, which can be immersed in a bath cryostat at TU Wien, Austria. The sample was contacted with thin gold wires in a four-probe configuration using a spot-welding device. It was then mounted on a sample puck with GE Varnish as the adhesive, and the probe was inserted directly into a bath cryostat, which is barely filled with liquid He. As the cryostat warms up, owing to thermal coupling to the environment, measurements are taken continuously whenever the temperature changes by approximately 1\,K. This is done by making use of an a.c. resistance bride (LakeShore 370).

The low-temperature Seebeck coefficient was measured with a separate custom-made setup at TU Wien, Austria. The tempeature differences and voltages are measured by making use of two chromel\,--\,constantan (\ce{Ni_{0.9}Cr_{0.1}}\,--\,\ce{Ni_{0.45}Cu_{0.55}}) thermocouples, which are soldered to both ends of the sample. Additionally, two strain gauges with a resistance of approximately 120$\,\Omega$ serve as heaters, attached to the bottom of each sample end using GE Varnish. This configuration allows for "seesaw heating" \cite{resel1996thermopower}, enabling the cancellation of spurious voltage contributions by switching the temperature difference during each measurement. The measurements are conducted in an evacuated sample chamber, where helium exchange gas can be introduced to provide a thermal coupling to the cryogen.

The thermal conductivity at low temperatures was measured using a steady-state method with a custom-built sample probe in a flow cryostat at TU Wien, Austria. A heater was attached to the top surface of the sample using a thermally conductive epoxy resin (STYCAST 2850FT). Two bundles of copper wires were first tightly fixed to the sample, to each of which a thermocouple was then soldered. The bottom of the sample was mounted on a copper heat sink, and measurements were conducted in a high vacuum of approximately 10$^{-5}$\,mbar. 

Measurements of the Hall effect at $2<T<300\,$K were carried out in a commercially available setup (PPMS by Quantum Design) and from $300<T<580\,$K in a home-built setup. For the latter, magnetic fields reaching up to 10\,T can be achieved via a superconducting magnet and a closed cycle refrigerator, which is thermally decoupled from the sample chamber through a high vacuum (approximately 10$^{-5}$\,mbar). Since the signal for such metallic samples is quite small, we fabricated specimen with a reduced thickness by polishing the samples down to approximately 20\,$\mu$m (see fig.\,S3A), which was possible due to the excellent mechanical properties of these materials. Measurements were carried out making use of the van der Pauw technique (see fig.\,S3B). 

\vspace*{-0.3cm}
\section*{Low-temperature thermodynamic measurements}
\vspace*{-0.2cm}
\noindent The temperature-dependent specific heat and magnetic susceptibility (see Fig.\,4A and fig.\,S12) from down to 2\,K and up to 320\,K were measured in a commercially available setup (PPMS by Quantum Design) making use of a standard relaxation-type calorimetry technique and a vibrating sample magnetometer (VSM), respectively. 
\vspace*{-0.3cm}
\section*{Screening Calculations}
\vspace*{-0.2cm}
\noindent In this section, we describe in further detail the screening method and provide the specific computational parameters relevant to the respective calculations. In general, for screening calculations we obtained the locally-relaxed structure, the total energy, and the density of states by means of Density Functional Theory as implemented in the Vienna Ab Initio Simulation Package (VASP) \cite{Kresse_PRB_1996_VASP}. We employed Projector-Augmented Wave (PAW) pseudopotentials provided in VASP \cite{Kresse_PRB_1999_VASP_pseudos} and the Perdew-Burke-Ernzerhof corrected for solids (PBEsol) exchange-correlation functional \cite{Perdew_PRL_2008_PBEsol}. Further details specific of each calculation are given below.

The first screening step was performed by fixing the crystal structure to a simple cubic unit cell with $AB_{3}$ composition. The purpose of this step is not to identify directly the correct structure, but rather to find pairs of atoms for which the relative chemical potentials are in the correct ballpark to place the strongly localized transition metal $3d$ states in the vicinity of the Fermi energy. Note that in binary transition metal alloys, where Ni atoms are embedded in face-centered-cubic Cu or Au lattices, the optimal concentration for the highest Seebeck coefficient is around \ce{Ni_{0.45}Cu_{0.55}} and \ce{Ni_{0.43}Au_{0.63}}. The power factor is even maximized at Ni-poorer compositions, for instance, \ce{Ni_{0.1}Au_{0.9}}. This suggests that an initial starting guess of $AB_{3}$ is not unreasonable, although future screenings with $AB$ or $AB_3$ might be worthwhile to investigate. The rationale is that if the chemical potentials are too different, there is no way at all in which the two atoms can combine in the desired way, and there is no point in looking further into potentially stable compositions. Within the cubic unit cell, the atoms $A$ and $B$ occupy the $1a$ and $3c$ Wyckoff positions, respectively. Using this structure as a template we proceeded to generate all possible combinations of the following atoms:
\begin{itemize}
	\item $A$ site: Fe, Co, Ni, Cu, Zn
	\item $B$ site: Li, Be, B, C, N, O, F, Na, Mg, Al, Si, P, S, Cl, K, Ca, Sc, Ti, V, Cr, Mn, Fe, Co, Ni, Cu, Zn, Ga, Ge, As, Se, Br, Rb, Sr, Y, Zr, Nb, Mo, Tc, Ru, Rh, Pd, Ag, Cd, In, Sn, Sb, Te, I, Cs, Ba, La
\end{itemize}

For each combination, we relaxed the crystal structure and computed the density of states (DOS) using four DFT steps, employing the output of one calculation as input for the successive one. In the following, we provide a summary of the computational parameters for these DFT calculations. 

\begin{enumerate}
	\item Structural relaxation at fixed volume (energy cutoff on plane waves: pseudopotential default; smearing 0.40 eV; k-points density: 0.30\,\AA$^{-1}$)
	\item Structural relaxation at variable volume (energy cutoff on plane waves: 600 eV; smearing 0.30 eV; k-points density: 0.30\,\AA$^{-1}$)
	\item Self-consistent calculation at the equilibrium volume (energy cutoff on plane waves: 600; smearing 0.20 eV; k-points density: 0.25\,\AA$^{-1}$)
	\item Non self-consistent calculation on a dense k-mesh (energy cutoff on plane waves: 600; smearing 0.20 eV; k-points density: 0.20\,\AA$^{-1}$)
	\item Postprocessing and extraction of Density of States
\end{enumerate}

An example of the result from the first screening for Ni as $A$ is shown in fig.\,S4. We note that although Ni$_1$Ge$_3$, shown in fig.\,S4, is not the structure investigated in the paper (which is Ni$_3$Ge$_1$), the sharp peak of the Ni-$3d$ states is nonetheless present about 1\,eV below the Fermi energy.

From the initial step, based on the position of the sharp $3d$ transition metal states relative to the Fermi energy, we established that the following pairs of atoms were promising: Mn-Mg, Fe-Ca, Fe-Zn, Co-Cd, Co-Cs, Co-Ga, Co-Mg, Co-Sn, Co-Zn, Co-Pb, Co-Hg, Ni-Hg, Ni-B, Ni-Ca, Ni-Ge, Ni-In, Ni-K, Ni-Rb, Ni-Sb, Ni-Si, Ni-Sn, Ni-Pb, Cu-Cs, Cu-Sc.

Starting from the pool of pairs of atoms identified in step 1, we performed crystal structure prediction calculations. To this end, we employed the variable-composition evolutionary algorithm implemented in the USPEX code \cite{Oganov_JCP_2006_uspex, Oganov_ACR_2011_evolutionary}.

For these calculations, we employed a population size of 80 individuals for the first generations, and 40 individuals for all the subsequent ones. Each calculation was run until the same structures were found on the convex hull for 4 generations, or for a maximum of 20 generations. Each structure underwent a six-steps relaxation with progressively tighter constraints, until forces on atoms were lower than 1\,e$^{-2}$ eV/\AA. A summary of all the convex hulls is shown in fig.\,S5 and an enlargement for the Ni-Ge system is shown in fig.\,S6. Each point corresponds to a crystal structure, each with its own energy. Compositions on the convex hull are stable with respect to any other possible decomposition. 

In Table\,\ref{tab:compositions}, we summarize all binary compositions which appear as thermodynamically stable in our structure searches. The respective crystal structures can be found in the crystallographic information file (CIF) format as accompanying data in the form of a compressed file.

\begin{table}[ht]
	\centering
\begin{tabular}{|c|c|}
\hline
 Element pair       &    Stable compositions \\
 \hline
 \hline
 Co-Cd                 &             None                  \\
  Co-Cs                 &             None                  \\
  Co-Ga                 &             Co$_3$Ga$_4$, CoGa, CoGa$_2$                  \\
  Co-Mg                 &             None                  \\
  Co-Sb                 &             CoSn$_2$                  \\
  Co-Zn                 &             CoZn$_5$, Co$_2$Zn$_3$, CoZn$_3$                  \\
  Cu-Cs                 &             None                  \\
  Cu-Sc                 &             Cu$_2$Sc, Cu$_5$Sc, Cu$_4$Sc, CuSc                  \\
  Fe-Ca                 &             None                  \\
  Fe-Zn                 &             None                  \\
  Co-Hg                 &             None                  \\
  Ni-Hg                 &             None                  \\
  Mn-Mg                 &             None                  \\
  Ni-B                 &             Ni$_4$B, Ni$_2$B, NiB, NiB$_8$         \\
  Ni-Ca                 &           NiCa$_3$, NiCa$_2$, NiCa,  Ni$_2$Ca, Ni$_5$Ca                  \\
  Ni-Ge                 &          Ni$_3$Ge, NiGe                  \\
  Ni-In                 &           NiIn                  \\
  Ni-K                 &             None                  \\
  Ni-Rb                 &             None                  \\
  Ni-Sb                 &             Ni$_6$Sb, Ni$_5$Sb, Ni$_3$Sb, NiSb                  \\
  Ni-Si                 &            Ni$_3$Si, Ni$_4$Si$_3$. Ni$_2$Si, NiSi$_2$                  \\
  Ni-Sn                 &            Ni$_3$Sn, Ni$_5$Sn$_3$, Ni$_3$Sn$_4$, NiSn                  \\
  Co-Pb                &            None                 \\
  Ni-Pb                 &            None                  \\
 \hline
\end{tabular}
\caption{Summary of compositions for which we found thermodynamically stable structures. The corresponding crystal structures are provided in the CIF format with the accompanying data.}
\label{tab:compositions}
\end{table}
\vspace*{-0.3cm}
\section*{Electron and Phonon Structure Calculations}
\vspace*{-0.2cm}
\noindent Electronic structure and phonon and electron-phonon calculations were performed using Quantum ESPRESSO \cite{Giannozzi_JPCM_2009_qe, Giannozzi_JPCM_2017_qe}, after re-relaxing the crystal structure appropriately. 

In particular, we employed Optimized Norm-Conserving Vanderbilt pseudopotentials \cite{Hamann_PRB_2013_ONCV}, with a cutoff of 80 Ry for the plane waves expansion of the Kohn-Sham wavefunctions. The ground-state charge density was obtained using a Monkhorst-Pack mesh of 16$\times$16$\times$16 and a smearing of 0.02\,Ry. We note that this mesh is denser than what is typically required with structures of similar volume. Due to the sharp features of the Fermi surface of this structure, however, this was required to achieve a convergence of the total energy to about 1\,meV/atom. We found the equilibrium volume to be especially sensitive to this convergence.

Phonon dispersions, densities of states and electron-phonon coupling (see fig.\,S7) were computed from linear response theory in the framework of Density Functional Perturbation Theory (DFPT) \cite{savrasov1994linear,baroni2001phonons}. In particular, the Eliashberg function $\alpha^2F(\omega)$, that is, the electron-phonon spectral function was obtained from the double-delta integral of the matrix element over the Fermi surface, as in eq. (18) of Ref. \cite{savrasov1994linear}.

From the Eliashberg function, we can obtain the cumulative integral of the electron-phonon coupling coefficient $\lambda_\text{e-p}(\omega)$ as the first inverse moment

\begin{equation}
\label{eq:lambda}
\lambda_\text{e-p}(\omega) = 2\int_{0}^{\omega_\text{max}}  \frac{\alpha^2F(\omega')}{\omega'} d\omega'
\end{equation}

From which we also have the total value $\lambda_\text{e-p} = \lambda_\text{e-p}(\omega_\text{max})$. We note, however, that $\lambda_\text{e-p}$ obtained this way, interestingly, does not account for the enhancement of the Sommerfeld coefficient of the specific heat (Fig.\,4A), suggesting that many-body effects beyond DFT and DFPT might become important when $E_\text{F}$ is placed in the vicinity of the flat band edge. 

The bands, Fermi surface and DOS were obtained via non-self consistent calculations. For the Fermi surface and DOS we employed a 36$\times$36$\times$36 mesh in reciprocal space. Phonon properties were computed from the converged self-consistent charge density, using a 6$\times$6$\times$6 grid for the phonon momenta. In addition, electron-phonon properties were computed by integrating the electronic states over a 24$\times$24$\times$24 grid, with a smearing of 200\,meV. The phonon dispersion alongside the atom-projected phonon density of states and Eliashberg function of \ce{Ni3Ge} are displayed in fig.\,S7.

\vspace*{-0.3cm}
\section*{Electronic Transport Calculations}
\vspace*{-0.2cm}
\noindent Transport calculations were performed using the Phoebe code~\cite{cepellotti2022phoebe}, an open-source package for electron and phonon Boltzmann transport equation solutions. Transport calculations were performed using the relaxation time approximation (RTA). The constant relaxation time approximation (CRTA) result was also calculated as a point of comparison. 
\\
Electron-phonon coupling calculations were performed using the JDFTx code~\cite{sundararaman2017jdftx}, with associated DFT calculations with the GBRV pseudopotentials~\cite{garrity2014pseudopotentials} parameterized for the PBE exchange-correlation functional~\cite{perdew1996generalized}. 

Here, electron phonon scattering rates were calculated using Wannier interpolation of the electron-phonon matrix elements, as described in~\cite{cepellotti2022phoebe}. The scattering rates were calculated as, 
%\begin{equation}
\begin{multline}
\frac{1}{\tau_{\boldsymbol{k}m}} = \frac{2 \pi}{V N_k} \sum_{m^{\prime} \boldsymbol{k}^{\prime}, \nu \boldsymbol{q}}\left|g_{m,m^{\prime},\nu}\left(\boldsymbol{k}, \boldsymbol{k}^{\prime}\right)\right|^2 
\\\times \big[ \left(1-f_{\boldsymbol{k}^{\prime} m^{\prime}}+n_{\boldsymbol{q} \nu}\right) \delta\left(\epsilon_{\boldsymbol{k} m}-\epsilon_{\boldsymbol{k}^{\prime} m^{\prime}}-\hbar \omega_{\boldsymbol{q} \nu}\right)  
\\ 
+\left(f_{\boldsymbol{k}^{\prime} m^{\prime}}+n_{\boldsymbol{q} \nu}\right) \delta\left(\epsilon_{\boldsymbol{k} m}-\epsilon_{\boldsymbol{k}^{\prime} m^{\prime}}+\hbar \omega_{\boldsymbol{q} \nu}\right) \big] \delta\left(\boldsymbol{k}-\boldsymbol{k}^{\prime}+\boldsymbol{q}\right) 
%\end{aligned}
\end{multline}
where $m$ is a band index, $\nu$ is a phonon mode index, $\boldsymbol{q}$ and $\boldsymbol{k}$ are phonon and electron wavevectors, with $\boldsymbol{k}' = \boldsymbol{k} + \boldsymbol{q}$, $\omega$ and $\epsilon$ are phonon and electron energies, and $n$ and $f$ are Bose-Einstein and Fermi-Dirac occupation factors respectively. From these lifetimes we predicted RTA-level transport properties as, 
\begin{equation}
[\rho]^{-1}_{\alpha,\beta}=\frac{e^2}{V N_k} \sum_{\boldsymbol{k}m} \frac{\partial f_{\boldsymbol{k}m}}{\partial \epsilon_{\boldsymbol{k}m}} \left(\boldsymbol{v}_{\alpha} \boldsymbol{v}_{\beta} \right)_{\boldsymbol{k}m} \tau_{\boldsymbol{k}m}
\end{equation}
for $\rho$, the electrical resistivity tensor and, 
\begin{equation}
[\sigma S]_{\alpha,\beta} = \frac{-e}{T V N_k} \sum_{\boldsymbol{k}m} \frac{\partial f_{\boldsymbol{k}m}}{\partial \epsilon_{\boldsymbol{k}m} } \left(\epsilon_{\boldsymbol{k}m} - \mu\right)\left(\boldsymbol{v}_{\alpha} \boldsymbol{v}_{\beta} \right)_{\boldsymbol{k}m} \tau_{\boldsymbol{k}m}
\end{equation}
to calculate $S$, where one can apply the inverse of the conductivity tensor calculated as above to isolate $S$. For constant relaxation time approximations, the expressions are identical, with the exception of $\tau_{km}\rightarrow\tau_{\mathrm{constant}} = 10\,$fs. Transport calculations were performed with 75$times$75$\times$75 k-point sampling for the integration of the Brillouin zone, adaptive smearing to approximate the Dirac delta functions, and the electronic states contributing to the calculation were selected by a threshold on occupation, using a Fermi window of width $\partial f / \partial T = 1 \times 10^{-10}$.

\vspace*{-0.3cm}
%\bibliographystyle{apsrev4-2}
%\bibliography{bibliography_Ni3Ge}

%\end{linenumbers}
\end{document}